# Intrinsic Temporal Coherence Governs Heat Transport of Zone-Folded Phonons

Xiaoyu Huang[1,2], Yuxiang Ni[1], Zhongwei Zhang[4], Yangyu Guo[5],
Marc Bescond[6], Masahiro Nomura[2,*], Sebastian Volz[3,4,*]

[1]School of Physical Science and Technology, Southwest Jiaotong University, Chengdu 610031, China
[2]Institute of Industrial Science, The University of Tokyo, Tokyo 153-8505, Japan
[3]Laboratoire des Solides Irradiés, Ecole Polytechnique, UMR CNRS 7642, CEA/DRF/IRAMIS,
28 route de Saclay, 91128 Palaiseau Cedex, France
[4]Center for Phononics and Thermal Energy Science, School of Physics Science and Engineering and China-EU Joint
Lab for Nanophononics, Tongji University, Shanghai 200092, China
[5]School of Energy Science and Engineering, Harbin Institute of Technology, Harbin 150001, China
[6]IM2NP, UMR CNRS 7334, Aix-Marseille Université, Marseille, France

While spatial phonon coherence manifested through band folding is believed to be a key factor governing the anomalous thermal conductivity of periodic structures, we investigate phonon transport from the perspective of temporal coherence. Using mode-resolved analyses, we quantify temporal coherent contributions and elucidate the interplay between phonon coherence time and lifetime in heat conduction of graphene/hexagonal boron nitride superlattices. We find that intrinsic coherence of folded phonon modes dominates the enhancement in ultrashort-period superlattices. In contrast, Wigner transport equation yields only a minor effect of band folding on thermal conductivity. The predictions in temperature dependence of models with and without temporal coherence provide a falsifiable experimental signature of this effect. Temporal coherence therefore constitutes a previously overlooked but fundamental channel for heat conduction, extending the conventional picture of spatially coherent transport and deepening the understanding of phonon dynamics in superlattices.

## I. INTRODUCTION.

Phononic crystals, in which a secondary periodicity is introduced, provide a powerful platform for controlling phonon propagation through dispersion engineering and wave interference. By tuning characteristic length scales such as period length, these systems can strongly modify heat transport and have therefore attracted broad interest in thermoelectric [1–3] and nanoscale thermal management [4,5].

Phonon heat transport in periodic nanostructures is commonly understood in terms of three spatial regimes: coherent (wave-like), incoherent (particle-like) and mixed [6–11]. A minimum in thermal conductivity is widely observed as the superlattice period decreases, signaling a crossover from incoherent to coherent transport. This behavior has been observed in various material systems both experimentally and numerically [6,8,12–15], and is usually interpreted within spatial coherence based on band folding. Such trends can be qualitatively reproduced by the Boltzmann transport equation (BTE) using folded phonon branches [16,17] and suggest new opportunities to tune thermal conductivity through structural design or temperature variation. For example, interface aperiodicity [18–20] or nanoscale obstacles [21,22] can induce destructive interference or even Anderson localization, thereby suppressing thermal conductivity. However, near the crossover regime, where bulk phonon modes mix with superlattice modes and diverse scattering processes occur, the microscopic origin of coherent contribution remain unsettled. In particular, existing interpretations focus mainly on spatial coherence associated with band folding, whereas the temporal persistence of phase correlations has received much less attention.

Much methodological progress has been made to model coherent phonon heat transport, yet different approaches capture coherence in fundamentally different ways. BTE-based models [14,23] can incorporate band folding and scattering through lifetimes, but treat transport primarily in the particle-like picture and do not provide a direct measure of phase persistence. Atomistic non-equilibrium Green's function (NEGF) [24–26] natively captures phonon wave effects, but incorporating realistic temperature-induced anharmonic scattering [27,28] remains computationally prohibitive for large systems like superlattices. Molecular dynamics (MD) simulations, widely used to capture both wave- and particle-like phonon behaviours, have demonstrated coherent transport in GaAs/AlAs [29] and Si/Ge [30,31] superlattices, as well as in two-dimensional nano-phononic crystals such as graphene-based systems [32,33]. However, MD alone does not provide a mode-resolved coherence diagnostic. What is missing is a mode-level link between structural periodicity and phonon phase persistence under scattering. Zhang et al. [34–36] provide such a framework by explicitly incorporating temporal coherence, motivating the analysis developed in this work.

Despite substantial progress, coherence in phonon heat transport is still often inferred indirectly from the overall trends or spectral decompositions. A more complete picture requires mode-resolved access to both spatial phase correlation and temporal phase persistence under varying thermal and structural conditions. Here we study coherent phonon heat transport in two-dimensional (2D) graphene/hexagonal boron nitride (h-BN) monolayer superlattices, which provide a clean model platform for examining the crossover from incoherent to coherent transport. We employ the formalism of Zhang *et al.* [35] to separate coherent and incoherent contributions, and cross-validate the results by complementary approaches. To connect periodicity to coherence, we analyze Brillouin-zone folding and use the Wigner transport equation (WTE) [37] to assess interbranch coherence associated with mode coupling. Crucially, we introduce a temporal criterion at the mode level by combining the coherence time $\tau^c$ and lifetime $\tau^p$ into the dimensionless ratio $\tau^c/\tau^p$, which quantifies coherence persistence relative to scattering. This framework introduces a mode-level temporal criterion to diagnose coherence. Although demonstrated here in 2D superlattices, the wave packet picture of coherent transport is general and should apply broadly to periodic nanostructures. It also predicts a near-plateau in the temperature dependence of thermal conductivity for short-period superlattices, offering a direct route to experimental tests near the crossover.

## II. METHODS

### A. Molecular Dynamics Simulation

All MD simulations were performed using the GPUMD package [38] with the Tersoff empirical potential [39,40]. The total dimensions of the lateral superlattice is 128 nm×10 nm [19], and periodic boundary conditions along all in-plane directions were imposed. At the initial stage of all simulations, the classical equations of motion were integrated with a 1 fs timestep. The systems were fully relaxed to zero pressure and room temperature in the isothermal-isobaric (NPT) ensemble for $3\times10^6$ timesteps (3 ns) using Berendsen barostat.

Equilibrium molecular dynamics (EMD) simulations were then mainly carried out to investigate the heat transport properties. The thermal conductivity was calculated using the following Green-Kubo formula [41–43] under a $1\times10^7$ timesteps (10 ns) microcanonical (NVE) ensemble simulation.

$$\kappa = \frac{V}{3k_BT^2}\int \langle J(t)\cdot J(0)\rangle dt. \quad (1)$$

where, $V$ is the system volume, $k_B$ is the Boltzmann constant and $T$ is the temperature. During this process, the cross-interface heat flux vector $J(t)$ was collected every 5 timesteps, autocorrelated over $2\times10^5$ timesteps (1 ns) and averaged over 10 sampling points before output. Each case was performed over 30 independent runs to obtain ensemble averaged values and the corresponding standard error of the mean (SEM). The proper decay of the heat flux autocorrelation function and the good convergence of the averaged thermal conductivity both confirmed the reliability of the simulations. In addition, EMD simulations with a 0.5 fs timestep were carried out to record atomic trajectories for subsequent wavelet transform analysis, that were sampled every 20 timesteps (0.1 ps) during a total $2\times10^5$ timesteps (1 ns) under the NVE ensemble.

### B. Lattice Dynamic Calculation

Both harmonic and anharmonic (third-order) vibrational properties were calculated based on the same Tersoff potential. A vacuum region of 10 Å was introduced along the out-of-plane direction to eliminate interactions between periodic layers. The harmonic phonon modes and dispersion relations were first obtained using Phonopy [44] with a 5×5×1 supercell and sampled with 51 ***k***-points along the selected high-symmetry path in the cross-interface direction. Subsequently, the anharmonic third-order interatomic force constants were calculated using Phono3py [44] with the same supercell size, where all considered cutoff distance is set to be 5 Å.

### C. Boltzmann Transport Equation Calculation

The lattice thermal conductivity of the superlattices was also calculated by solving BTE within the conventional single-mode relaxation time approximation (SMRTA), using the implementation in Phono3py [44]. To ensure accuracy, phonon lifetimes were evaluated with the tetrahedron method for Brillouin zone integration, and the sampling meshes were systematically converged for each system (Fig. S1).

### D. Coherent Heat Transport Calculation

Phonon coherence in heat transport is often discussed in a unified manner, yet it can arise from physically distinct mechanisms. Here we distinguish mutual coherence from intrinsic coherence at the operator level.

*Contact author: nomura@iis.u-tokyo.ac.jp; sebastian.volz@cnrs.fr

In the WTE formalism [37], transport is described by the phonon density matrix $\rho_{ss'}(\mathbf{k})$ in the harmonic eigenmode basis. Diagonal elements $\rho_{ss}(\mathbf{k})$ give mode populations, whereas off-diagonal elements $\rho_{ss'}(\mathbf{k})(s \neq s')$ encode coherence between different branches $s(s')$ at the same wave vector $\mathbf{k}$. The thermal conductivity can be written schematically as a sum of population $\kappa_p$ and coherence terms $\kappa_c$, where $\kappa_p$ is the lattice thermal conductivity from BTE. $\kappa_c$ arises from off-diagonal velocity matrix elements $v(\mathbf{k})_{ss'}$ coupled to $\rho_{ss'}(\mathbf{k})$. In the steady-state regime, $\rho_{ss'}(\mathbf{k})$ can be solved analytically and substituted back, yielding the following expression:

$$\kappa_c = \frac{\hbar^2}{k_B T^2} \frac{1}{V N_c} \sum_{\mathbf{k}} \sum_{s \neq s'} \frac{\omega(\mathbf{k})_s + \omega(\mathbf{k})_{s'}}{2} v(\mathbf{k})_{ss'} v(\mathbf{k})_{s's}$$

$$\times \frac{\omega(\mathbf{k})_s \bar{N}(\mathbf{k})_s (\bar{N}(\mathbf{k})_s + 1) + \omega(\mathbf{k})_{s'} \bar{N}(\mathbf{k})_{s'} (\bar{N}(\mathbf{k})_{s'} + 1)}{4(\omega(\mathbf{k})_s - \omega(\mathbf{k})_{s'})^2 + (\Gamma(\mathbf{k})_s - \Gamma(\mathbf{k})_{s'})^2}$$

$$\times (\Gamma(\mathbf{k})_s + \Gamma(\mathbf{k})_{s'}). \quad (2)$$

where, $V$ is the unit cell volume, $\bar{N}(\mathbf{k})_s = (e^{\hbar\omega(\mathbf{k})_s/k_B T} - 1)^{-1}$ is the equilibrium Bose-Einstein distribution, and $\Gamma(\mathbf{k})_s$ is the phonon linewidth extracted from Lorentzian fitting of spectral energy density (SED). $v(\mathbf{k})_{ss'}$ is obtained from the product of the phonon eigenvectors $\mathbf{e}(\mathbf{k})$ and the derivative of the harmonic dynamical matrix $D(\mathbf{k})$ with respect to the wave vector $\mathbf{k}$. This interbranch coherence becomes appreciable when branch frequency separation is comparable to the anharmonic linewidths. Physically, it describes mutual coherence mediated by mode coupling enabled by spectral overlap [45,46].

We also employed the generalized heat conduction theory of Zhang *et al* [34,35]. In this approach, phonons are treated as wave packets constructed within a single branch $s$. We projected the dynamics information onto individual modes in the time-frequency domain via wavelet transform. From the decay of the wave packet envelope we extracted two timescales, lifetime $\tau^p$ and coherence time $\tau^c$ (see Supplementary Methods). These timescales need not coincide. While $\tau^p$ characterizes energy relaxation, $\tau^c$ quantifies how long the phase of the wave packet remains correlated. Because the analysis is based on MD trajectories, the extracted $\tau^c$ reflects the combined impact of anharmonicity and lattice dynamics on phase persistence. It therefore serves as an effective diagnostic of temporal coherence, reflecting modes coupling and intrinsic coherence, the natural temporal and spatial extent of a phonon wave packet. Importantly, this intrinsic phase persistence is intraband and does not rely on off-diagonal density matrix elements between branches. It can remain significant even when interbranch coupling is weak. In this framework, the classical particle-like thermal conductivity reduces to the standard BTE expression:

$$\kappa_p = \frac{1}{3} \sum_\alpha \sum_{ks} C_{v,ks}^{clas} v_{ks,\alpha}^2 \tau_{ks}^p. \quad (3)$$

The complete thermal conductivity ($\kappa_{p+w}$) including both the particle-like ($\kappa_p$) and wave-like ($\kappa_w$) contributions is:

$$\kappa_{p+w} = \frac{1}{3} \sum_\alpha \sum_{ks} C_{v,ks}^{clas} v_{ks,\alpha}^2 \sqrt{\frac{\pi}{4ln2}} \tau_{ks}^c e^{\frac{{\tau_{ks}^c}^2}{128 ln2 {\tau_{ks}^p}^2}}. \quad (4)$$

where, $C_{v,ks}^{clas} = k_B/V$ is the classical specific heat per mode. A quantum correction can be applied to compare with that from WTE by scaling the classical mode contribution with the heat capacity ratio $\alpha = C_{v,ks}^{quan}/C_{v,ks}^{clas} = x^2 e^x/(e^x - 1)^2$, where $x = (\hbar\omega_{ks}/k_B T)$. $v_{ks,\alpha}$ denotes the group velocity of the specific phonon mode along the Cartesian coordinate $\alpha$, which in this study corresponds to the cross-interface direction. When the mode $\tau_{ks}^c$ equals $\tau_{ks}^p$, this expression naturally reduces to the conventional BTE with a mirror correction factor (≈1.07).

Both Eq. (2) and Eq. (4) are mode-level approaches based on phonon dispersion relations (see Fig. S2 for a comparison between mode-resolved and spectral-resolved decompositions). Here, Eq. (2) was evaluated using Phono3py, and Eq. (4) was implemented in the WPPT package [47]. To reduce computational cost, we uniformly sampled 11 $\mathbf{k}$-points along the transport direction and evaluated over all vibrational modes at these points. Considering the anharmonic scattering and broadening spectrum from temperature effect, the phonon dispersion calculated from Phonopy (0 K) was compared with the SED analysis at 300 K. Their excellent agreement (Fig. S3) supports the use of harmonic dispersions for the coherent heat transport simulations at room temperature.

### III. RESULTS AND DISCUSSION

#### A. Thermal conductivity at room temperature

Graphene has weak Umklapp scattering, and graphene/h-BN interfaces can be atomically smooth because of their nearly matched lattice constants. These features promote specular phonon reflection and make 2D graphene/h-BN monolayer superlattices a suitable model system for studying coherent phonon heat transport. Although such monolayer superlattices have not yet been realized experimentally, coherent phonon transport is expected in superlattices with short periods, high quality interfaces, and weak anharmonicity. The underlying physics is not restricted to a particular material or dimensionality. The superlattice models were constructed using the zigzag interface configuration [48], with the total dimensions fixed at 128 nm×10 nm and the period lengths varied between $L_p = 0.85 \sim 4.27$ nm (see Fig. 1(a), orange box). This setup allows us to probe crossover from coherent to incoherent heat transport.

*Contact author: nomura@iis.u-tokyo.ac.jp;
sebastian.volz@cnrs.fr

We first compute the lattice thermal conductivity of graphene/h-BN superlattices at 300 K using MD simulations, which naturally includes phonon–phonon, phonon-interface scattering as well as coherence effects. However, its heat statistics are classical and are strictly valid in the Maxwell–Boltzmann high-temperature limit (that is $T \gg T_D$, where $T_D$ is the Debye temperature). As shown in Fig. 1(b), the classical thermal conductivity exhibits a non-monotonic dependence on $L_p$, initially decreasing and then increasing as $L_p$ becomes larger. This behaviour is often interpreted as a change in the dominant heat transport picture, from coherent/wave-like propagation to incoherent diffusive scattering. Here, coherence refers to spatial coherence associated with band folding. Four representative periods $L_p$ = 0.85, 1.71, 2.56 and 4.27 nm are selected for following detailed analysis.

Given the high Debye temperature of graphene and the broad frequency spectrum contributing to heat transport, quantum statistics must be accounted for quantitative comparison. Fig. 1(b) compares thermal conductivities obtained using different approaches and shows overall consistency. Specifically, the BTE results (upper triangles) including only particle-like phonons agree well with $\kappa_p^{quan}$. Their limited values at short periods already suggest a non-negligible coherent contribution. When temporal coherence is further included, the total thermal conductivity $\kappa_{p+w}^{clas}$ closely reproduces the MD results (grey region). Together, these agreements indicate that the generalized heat conduction theory in Eq. (4) captures both particle-like and wave-like phonon transport in this system. Moreover, the gap between the red and blue curves decreases with increasing $L_p$, indicating progressive suppression of phonon coherence and a crossover from combine spatially and temporally coherent transport to an incoherent regime.

The WTE has been proposed as an alternative description of phonon coherence in crystalline and amorphous solids. However, its predictions in our superlattices deviate from both Eq. (4) and MD results, instead collapsing toward the BTE limit. This would imply negligible coherence, which is physically unreasonable for short-period superlattices with high-quality interfaces. We discuss the origin of discrepancy below.

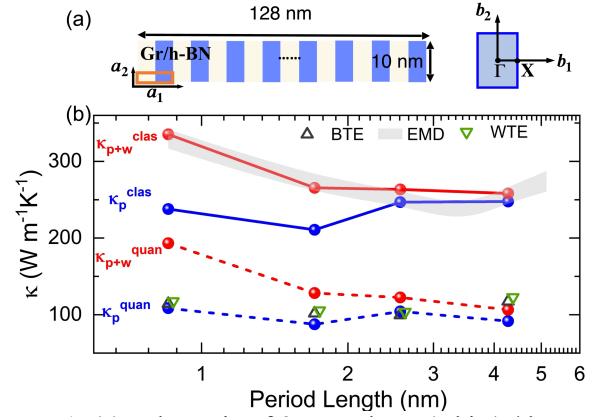

FIG 1. (a) Schematic of 2D graphene (white) / h-BN (blue) monolayer superlattices with period length $L_p$. The orange box marks the unit cell. The corresponding first Brillouin zone and selected high-symmetry points are shown on the right. (b) Thermal conductivity at 300 K as a function of period length $L_p$, obtained from different approaches. The grey region shows the MD results with SEM error bars, the upper and lower triangles denote the BTE and WTE results, respectively. Filled symbols represent the results from Eq. (3) and Eq. (4), evaluated in the classical limit (solid lines) and with quantum correction (dashed lines), for $L_p$ = 0.85, 1.71, 2.56 and 4.27 nm.

### B. Band folding and mutual coherence

This transport behaviour can be interpreted by tracking how the phonon dispersions evolve with $L_p$. Under Brillouin-zone folding [14], the superlattice dispersions can be viewed as folded versions of the parent dispersions of graphene and h-BN. Increasing $L_p$ enlarges the real space unit cell and correspondingly densifies the phonon branches in reciprocal space. This spectral crowding increases the chance of interbranch coupling. Accordingly, Fig. 2(a) shows a larger mutual coherence fraction estimated as $[(\kappa_{WTE} - \kappa_{BTE})/\kappa_{WTE}] \times 100\%$. However, this coherence is mediated by anharmonic interactions. It becomes more pronounced at elevated temperatures shown in Fig. S4, effectively a form of mutual coherence. Although the mutual coherence increases with $L_p$, it remains small in magnitude. This suggests that mutual coherence is not the dominant mechanism behind the large thermal conductivity changes.

*Contact author: nomura@iis.u-tokyo.ac.jp; sebastian.volz@cnrs.fr

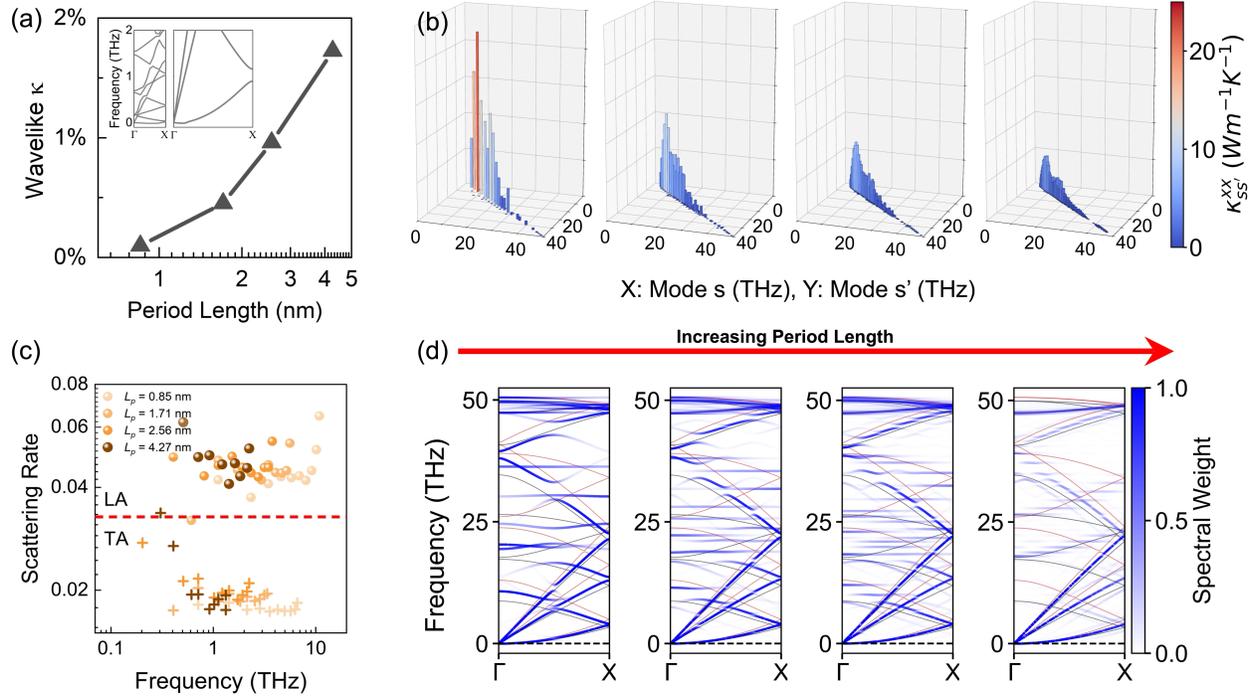

FIG 2. Band folding and mutual coherence in 2D graphene/h-BN superlattices with $L_p$ = 0.85, 1.71, 2.56 and 4.27 nm at 300 K. (a) Mutual coherence contribution to thermal conductivity estimated from WTE as a function of $L_p$. The inset shows phonon dispersion of $L_p$ = 0.85 and 4.27 nm. (b) Mode-resolved cross interface thermal conductivity distributions obtained from WTE. The colour scale indicates the magnitude of the mode contribution. Each pixel represents coupling between a pair of phonon modes. The diagonal features correspond to couplings between quasi-degenerate states and are mainly associated with particle-like transport. The off-diagonal features reflect interbranch coherent couplings between modes with distinct frequencies, which can be interpreted as wave-like tunnelling processes. (c) Scattering rates of LA and TA modes. (d) Unfolded phonon dispersion maps for each period. The colour indicates the projection weight, and brighter colours imply stronger Bloch character. Red lines are the phonon dispersions of graphene with rectangular primitive cell, and grey lines are those of h-BN.

To examine how band folding redistributes heat carrying modes, Fig. 2(b) presents the mode-resolved thermal conductivity map from WTE. WTE decomposes the phonon heat transport into population (diagonal) and coherence (off-diagonal) contributions within a unified framework. The population term is dominated by low-frequency modes with large group velocities and weak anharmonic scattering, consistent with the particle-like picture captured by BTE. In contrast, the coherence term arises from interbranch couplings. For all $L_p$ considered, the dominant contribution comes from quasi-degenerate couplings near the diagonal. This helps explain why the WTE maps largely reproduce the BTE predictions in Fig. 1. Low-frequency acoustic modes contribute most, and the contributing range gradually extends to mid-high frequencies as $L_p$ increases. Even so, the total thermal conductivity increases only slightly within WTE.

To link band folding to phonon transport mechanisms, we analyze the results from both particle-like and wave-like perspectives. Fig. 2(c) compares the scattering rates of longitudinal acoustic (LA) and transverse acoustic (TA) modes. At longer $L_p$, low-frequency modes show higher scattering rates because more scattering channels become available. High-frequency modes that remain particle-like experience reduced scattering as $L_p$ increases. As a result, the particle-like contributions are redistributed across frequencies.

Furthermore, band unfolding is performed to probe spatial phase correlation across interfaces. In an ideal periodic crystal, phonons transport as Bloch waves, $\psi(\mathbf{r}) = A(\mathbf{r})e^{i\mathbf{k}\cdot\mathbf{r}}$, which can be viewed as plane waves whose amplitude is modulated by a lattice-periodic function $A(\mathbf{r})$. They exhibit a well-defined phase relation under lattice translations $\psi(\mathbf{r}+\mathbf{r}_j) = \psi(\mathbf{r})e^{i\mathbf{k}\cdot\mathbf{r}_j}$ and thus extend throughout a crystal are coherent in their lifetime. In superlattices, the modes remain Bloch waves with respect to the supercell periodicity, but their Bloch character with respect to the primitive cell can be redistributed among multiple wave vectors. Band unfolding quantifies this redistribution by projecting a supercell eigenmode $|ks\rangle$ onto the primitive cell Bloch states at $k + G_i$; the resulting projection weight $W_{ks}(G)$, that is Bloch character, measures how well primitive cell translational symmetry is preserved.

*Contact author: nomura@iis.u-tokyo.ac.jp;
sebastian.volz@cnrs.fr

Specifically, we project each calculated supercell eigenstates $|ks\rangle$ onto Bloch states of the rectangular primitive cell using the translation projection operator $\hat{T}(r_j)$, defined as $\hat{T}(r_j)f(r) = f(r + r_j)$. Following Allen et al. [49], the projection weight is:

$$W_{ks}(G) = \frac{1}{\mathcal{N}} \sum_{j=1}^{\mathcal{N}} \langle ks|\hat{T}(r_j)|ks\rangle e^{-i(k+G_i)\cdot r_j} \quad (4)$$

where, $r_j$ are the $\mathcal{N}$ distinct primitive cell translation vectors that generate the supercell and $G_i$ are the corresponding $\mathcal{N}$ distinct reciprocal lattice vectors that map supercell $k$ to primitive Brillouin zone $k + G_i$. The sum rule $\sum_i W_{ks}(G_i) = 1$ ensures that the total weight of a supercell mode is conserved and redistributed over primitive cell wave vectors. In Fig. 2(d), brighter colours indicate larger weights, i.e., stronger Bloch character and better preservation of primitive cell translational symmetry.

At short $L_p$, compositional modulation, such as mass and force constant, reduces the primitive translational symmetry. As a result, the unfolded spectra show clear folded replicas and these branches are not simply those of pristine graphene or h-BN. Their unfolding weights are high, meaning the modes keep strong Bloch character with respect to the reference primitive cell. This indicates better translational coherence over several periods and supports more wave-like transport. Consistent with this picture, the EMD thermal conductivity in the ultrashort-period reaches ~329 W/mK, compared with ~230 W/mK at longer periods (Fig. 1 (b)). As $L_p$ increases, modes become more confined within individual layers and the unfolded spectral weight progressively concentrates around features resembling those of the constituent materials. The low-frequency acoustic modes representing collective vibration remain robust throughout, with frequencies lying between those of graphene and h-BN. In contrast, mid-frequency modes that are distinct at ultrashort $L_p$ become diffuse at longer $L_p$. Overall, the Bloch character is redistributed over multiple wavevectors, and phase correlation is reduced. This trend is consistent with a crossover to incoherent, particle-like transport. In this work, $L_p \approx 4$ nm marks the crossover. Above this value, the population term dominates.

Clear coherence enhanced heat transport is mainly observed at short periods where folded branches with large unfolding weights persist.

### C. Intrinsic coherence and temporal characteristics

The heat transport in superlattices described by the WTE mainly captures the populations and the mutual coherence associated with interbranch couplings induced by band folding, and thus misses part of the coherent contribution. To clarify the origin of the coherent contribution and the transport crossover in Fig. 1, we examine the temporal characteristics of phonon transport using Eq. (4). In this framework, coherence is governed by the group velocity and two characteristic times, namely the coherence time $\tau^c$ and the lifetime $\tau^p$. Folded branches control phase correlation across interfaces and therefore spatial coherence. In contrast, the temporal behavior of each mode determines how long this correlation persists during transport. Accordingly, we characterize the dynamical persistence of coherence by the ratio $\tau^c/\tau^p$. A larger $\tau^c/\tau^p$ indicates stronger temporal coherence and thus greater potential for coherent heat transport.

Following the same idea as mutual coherence, we quantify the coherent contribution in Fig. 3(a) as $[(\kappa_{p+w}^{quan} - \kappa_{BTE})/\kappa_{p+w}^{quan}] \times 100\%$. This fraction decreases from ~45% to ~10% as $L_p$ increases, consistent with the crossover in Fig. 1. Unlike the mutual coherence in Fig. 2(a), this contribution is dominated by intrinsic coherence. Here it refers to phase correlation between plane waves within the same phonon branch. This intrinsic coherence is not represented by the anharmonic interbranch coupling terms emphasized in WTE, yet it provides the main coherent channel in these artificial superlattices. Additional support comes from the enhanced coherent fraction at lower temperatures (Fig. S5). Under weak anharmonicity, intrinsic coherence persists even when interbranch coupling is weak, highlighting a departure from the purely particle-like BTE picture.

As $L_p$ increases, Brillouin zone folding produces flatter branches near the zone boundary and slightly reduces the group velocity. This trend differs from earlier explanations that attributed the reduced thermal conductivity in the coherent regime partly to changes in group velocity caused by band folding [17]. Instead, $\tau^c/\tau^p$ serves as a practical indicator of coherent transport and offers a unified temporal interpretation for the minimum thermal conductivity in superlattices.

Fig. 3(b) shows mode-resolved heatmaps of $\tau^c/\tau^p$ for different $L_p$. Comparing these maps reveals that the LA branch, including folded counterparts, dominates the coherent contribution. Modes below ~25 THz exhibit both large $\tau^c/\tau^p$ and substantial group velocities, consistent with the accumulative thermal conductivity in Fig. 3(c) (see Fig. S6 for other $L_p$). As $L_p$ increases, coherent modes shift towards the Brillouin zone boundary, where flatter dispersions favor spatial coherence. The typical $\tau^c/\tau^p$ level remains nearly unchanged, suggesting a redistribution of coherent modes rather than a strong reduction in intrinsic coherence.

*Contact author: nomura@iis.u-tokyo.ac.jp;
sebastian.volz@cnrs.fr

After including the corresponding group velocities, Fig. 3(d) shows the resulting mode-resolved thermal conductivity. Although some zone boundary modes retain partial spatial and temporal coherence, their contribution to heat transport is small due to their low group velocities. As a result, the total thermal conductivity decreases at longer $L_p$. Together, the temporal analysis based on $\tau^c/\tau^p$ and the mode-resolved thermal conductivity build a microscopic picture consistent with the overall trend in Fig. 1, linking coherence variations to measurable heat transport.

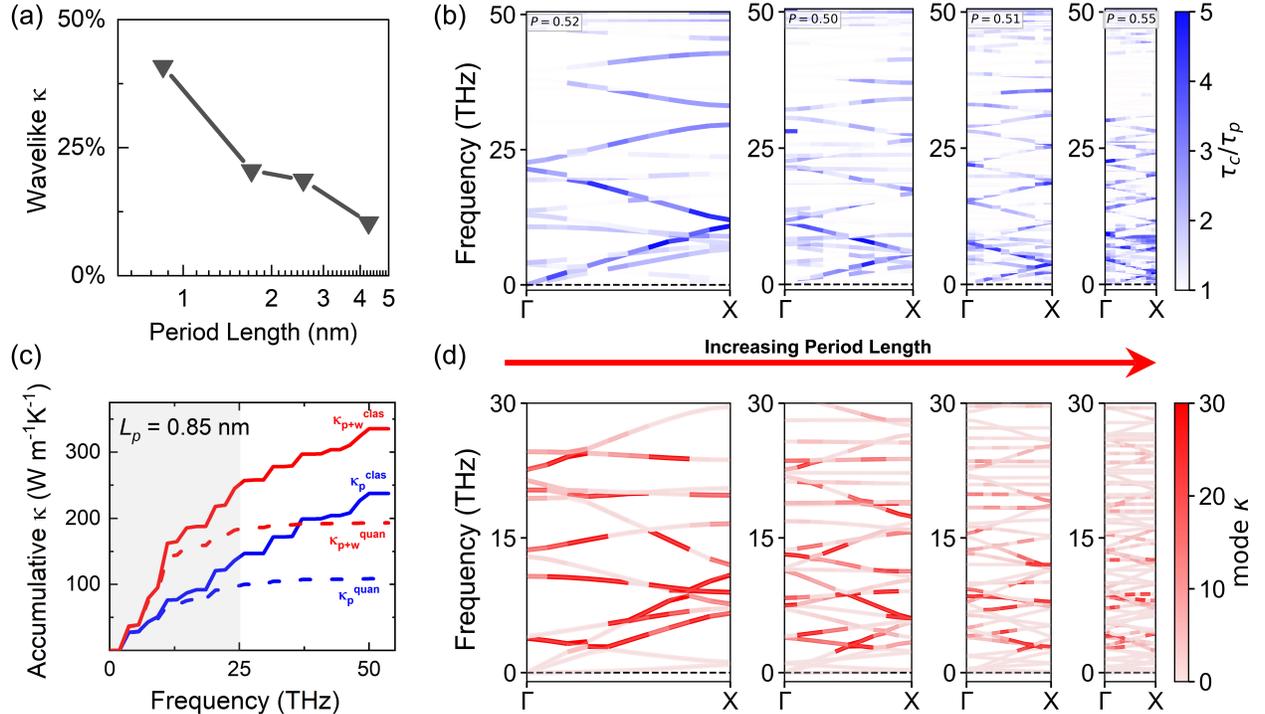

FIG 3. Intrinsic coherence and temporal characteristics of phonon modes in 2D graphene/h-BN superlattices. (a) Intrinsic coherence contribution to thermal conductivity estimated from Eq. (4) as a function of $L_p$. (b) Mode-resolved maps of $\tau^c/\tau^p$ for $L_p$ = 0.85, 1.71, 2.56 and 4.27 nm. Coloured regions indicate phonon modes included in the analysis. Regions with $\tau^c/\tau^p > 1$ correspond to modes that can participate in temporally coherent transport, and brighter colours indicate larger $\tau^c/\tau^p$. $P = \frac{1}{N}\sum_{i=1}^{N} \Pi(\tau_c/\tau_p > 1)$ denotes the fraction of modes with $\tau^c/\tau^p > 1$, where $\Pi$ is the indicator function (equal to 1 if the condition is satisfied and 0 otherwise). (c) Accumulative thermal conductivity for $L_p$ = 0.85 nm at 300K. (d) Mode-resolved thermal conductivity for $L_p$ = 0.85, 1.71, 2.56 and 4.27 nm, and brighter colours indicates larger contributions.

### D. Temperature dependence of thermal conductivity

So far, we have clarified the physical origin of the thermal conductivity minimum in these superlattices. We next examine the temperature dependence of thermal conductivity $\kappa(T)$, as shown in Fig. 4. The classical results exhibit the expected monotonic decrease with temperature, consistent with stronger anharmonic phonon scattering at higher $T$.

In contrast, the temperature dependence of quantum-corrected $\kappa^{quan}(T)$ depends strongly on the $L_p$. In the ultrashort-period, $\kappa_{p+w}^{quan}(T)$ decreases with temperature, consistent with transport dominated by intrinsic coherence. In this regime, phonon transport is largely wave-like, while the particle-like contribution $\kappa_p^{quan}(T)$ shows only weak temperature dependence (Fig. 4(d)). As $L_p$ increases but remains below ~4 nm, a growing fraction of modes becomes particle-like. The combined effects of increased thermal population and anharmonic scattering lead to a weak increase in the $\kappa_p^{quan}(T)$ at higher $T$. Meanwhile, the intrinsic coherence contribution decreases with temperature. The competition between these two trends give rise to a near-plateau in the total $\kappa_{p+w}^{quan}(T)$. This behaviour is consistent with experimentally reported weak temperature dependence in short-period superlattices [6].

*Contact author: nomura@iis.u-tokyo.ac.jp; sebastian.volz@cnrs.fr

Beyond the well-known minimum in $\kappa$ versus $L_p$, this temperature dependence provides an additional quantitative signature of coherent transport. A near-plateau in $\kappa(T)$ is expected only for short periods, whereas long-period structures should show a more conventional decrease with $T$. This yields a clear, falsifiable criterion that is experimentally accessible, for example via time-domain thermoreflectance measurements. Recent progress in aligned graphene/h-BN stitching provides a realistic route to test these predictions, including atomically sharp zigzag interfaces [48], controllable length scales [50], and centimeter-scale growth enabled by chemical vapor deposition [51].

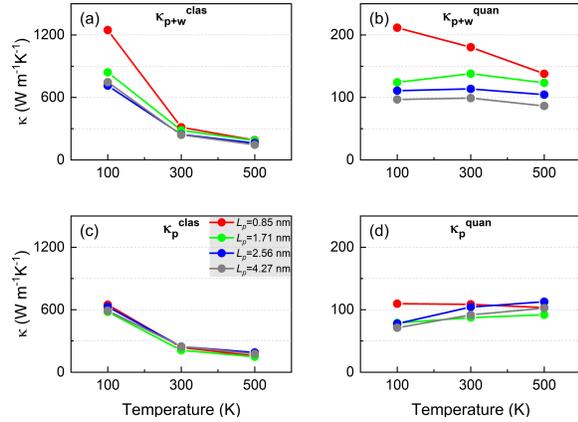

FIG 4. Temperature dependence of thermal conductivity in 2D graphene/h-BN superlattices with $L_p$ = 0.85, 1.71, 2.56 and 4.27 nm, represented by colours. Results are obtained from Eq. (3) and Eq. (4) in the classical limit (a)(c) and with quantum correction (b)(d).

## IV. CONCLUSIONS

In this work, we investigate coherent phonon heat transport in 2D graphene/h-BN monolayer superlattices using mode-resolved analyses. The interplay between structural periodicity and phonon coherence leads to a nonmonotonic dependence of thermal conductivity on period length $L_p$, with a crossover at $L_p \approx 4$ nm from spatially coherent to incoherent transport. We clarify how structural periodicity shapes the coherent contribution. Brillouin-zone folding enhances spectral crowding, which can promote interbranch coupling and a small mutual coherence component. Band unfolding provides direct spatial evidence for coherence. Short-period structures retain strong Bloch character and clear folded branches with robust phase correlation across interfaces, whereas long-period superlattices exhibit diffuse unfolded weights and reduced translational coherence due to interfacial phase randomization and mode confinement. On the temporal side, we identify the temporal persistence of coherence as a key determinant of coherent heat transport. Within Eq. (4), intrinsic coherence dominates in short-period superlattices. The ratio $\tau^c/\tau^p$ provides a compact indicator of this behavior. Mode-resolved maps show that the longitudinal acoustic branch contributes most strongly, with the dominant coherent modes mainly below ~25 THz, where sizable group velocities and persistent phase correlation coexist. The temperature dependence further yields a testable signature of coherence. For short periods, the competition between a decreasing coherence contribution and an increasing particle-like contribution leads to a near-plateau in $\kappa(T)$. Together, these results link structural periodicity, spatial and temporal coherence, and the overall heat transport. They also provide practical guidelines for engineering coherence-enhanced heat conduction in low-dimensional heterostructures and related phononic architectures.

## ACKNOWLEDGMENTS

This project is supported by KAKENHI (grant No. 21H04635) and the ANR project COCONUT (ANR-24-CE50-5682).

All data needed to evaluate the conclusions in the paper are present in the paper and/or the Supplementary Materials. Additional data related to this paper may be requested from the authors.

See Supplementary Material for details on: (1) coherent heat transport calculation using wavelet transform (2) convergence test for BTE calculation, (3) comparison between mode and spectral decomposition, (4) comparison between harmonic phonon dispersion and SED at 300K, (5) the temperature dependence of coherence, and (6) accumulative thermal conductivity for other periods.

*Contact author: nomura@iis.u-tokyo.ac.jp; sebastian.volz@cnrs.fr

*Contact author: nomura@iis.u-tokyo.ac.jp;
sebastian.volz@cnrs.fr

*Contact author: nomura@iis.u-tokyo.ac.jp; sebastian.volz@cnrs.fr